\def\R{{I\!\! R}}

\documentstyle{article}
\begin{document}
\begin{titlepage}

\title{The Universality of Einstein Equations}

\author{Marco FERRARIS\\
Dipartimento di Matematica, Universit\`a di Cagliari\\
Via Ospedale 72, 09124 CAGLIARI (ITALY)
\and Mauro FRANCAVIGLIA\and Igor VOLOVICH\thanks
{Permanent address: Steklov Mathematical
Institute, Russian Academy of Sciences \goodbreak
Vavilov St. 42, GSP--1, 117966
MOSCOW (RUSSIA)}\\
Istituto di Fisica Matematica ``J.--L. Lagrange''\\
Universit\`a di Torino\\ Via C. Alberto 10, 10123 TORINO (ITALY)
}
\date{May 5, 1993}
\maketitle

\begin{abstract}

It is shown that for a wide class of
 analytic Lagrangians which
depend only on the scalar curvature of a
 metric and a connection,
the application of the
so--called ``Palatini formalism'', i.e., treating
 the metric and the
connection as independent variables, leads
to ``universal'' equations.
If the dimension $n$ of space--time is
greater than two these universal equations
are Einstein equations for a generic Lagrangian and are
 suitably replaced
by other universal equations at bifurcation points.
We show that bifurcations take
 place in particular
for conformally invariant
 Lagrangians $L=R^{n/2} \sqrt g$ and prove that their
solutions are conformally equivalent to solutions
 of Einstein equations.
For 2--dimensional space--time we find instead that
the universal equation is always the equation
of constant scalar curvature;
the connection in this case is a Weyl
 connection, containing the
Levi--Civita connection of the metric and an additional
 vectorfield ensuing
from conformal invariance.
As an example, we investigate in detail
 some polynomial
Lagrangians and discuss their bifurcations.

\end{abstract}
\end{titlepage}
\section{Introduction}

In this paper we shall investigate a large class of
 gravitational
Lagrangians depending on a metric and a torsionless
 connection as basic
dynamical variables. We shall show that Einstein equations
 play always a
very central role in the formalism, because of
 a strong ``universality''
property.

We first remark that, according to a widespread
 opinion, one is usually
led to believe that any physically meaningful
 set of field equations
having a variational character should be in
 correspondence with an
``essentially unique'' Lagrangian, in the sense
 that for any given
set of field variables the Lagrangian generating
the equations should be
fixed modulo the addition of (dynamically
irrelevant) divergencies.
However, this is not true, because one can produce
 examples of non--trivially
related families of Lagrangians which are ``strongly
 dynamically equivalent'',
in the sense that for a fixed set of field variables
 they generate the same
field equations, although they
do not differ by divergencies. There exist
other approaches to "dynamical equivalence" which
 allow instead to recognize
equivalence through suitable changes of variables, e.g. by
Legendre transformation
or exploiting invariance properties. In
gravitational theories
their early history begins with Einstein,
Eddington and Weyl (see
 e.g. \cite{Ein,Edd,Wey}), passes through
Schr\"odinger \cite{Sch} and still today is the
 subject of interest in
the literature (see e.g. \cite{FK,Fer,JK,MFF1}
and ref.s quoted therein).
In this paper we shall consider equivalence
in the ``strong'' sense above,
i.e., by assuming field variables to be fixed
 and not changing them.

This raises an important problem, which we
 believe should be given renewed
attention for better understanding both the
classical and the quantum
properties of physical fields, namely: `` which
is the most central
object in a field theory? Is it the set of
field equations or the
Lagrangian?''.

As we shall see below, Einstein equations are a
 strikingly important
example for the discussion of this problem. We shall
 in fact show and analyze
their ``universality'', by proving that a very large
class of profoundly
different metric--affine Lagrangians (i.e., depending
 on a metric and
a connection), which one should expect in
principle to generate rather
different equations, always produce the same set
 of ``universal'' field equations.
These will be in fact Einstein equations in
 generic cases. In degenerate
situations and in dimension $n=2$, one gets instead
equations which express the constancy
of the scalar curvature (which contain Einstein
equations as a sector) or,
for conformally invariant Lagrangians, equations which are
conformally equivalent to Einstein equations.

Since the early days of General Relativity, the
 so--called ``Palatini
formalism'', which is based on independent variations
 of the metric and
the connection,
has been known to be equivalent to the corresponding metric
formulation for the Lagrangian
which depends linearly
on the scalar curvature constructed out of the metric
and the Ricci tensor
of the connection. In this case, in fact, when $n\ge 2$
field equations imply that the connection has to be
the Levi--Civita
connection of the metric, which in turn satisfies Einstein
equations (see, e.g., \cite{Ray}). This means that there is
a full dynamical
equivalence with the corresponding metric Lagrangian, which a posteriori
coincides with the Hilbert
Lagrangian of General Relativity. (For a
historical discussion about
Palatini formalism see \cite{FFR}).

In this paper we will show that the method of
independent variations of the metric and the connection leads
in fact to Einstein equations
for a much wider class of Lagrangians
which depending on the scalar curvature in a
 non--linear way. In the
process of showing that the relevant field equations reduce to
Einstein equations
a cosmological constant $\Lambda$ arises, whose
only allowed values
are fully determined by the Lagrangian itself.
One thus finds that Einstein equations
are not strictly related with the
linear Lagrangian, since any other Lagrangian
 belonging to the family
can be used to derive them. This displays the
 claimed ``universality'' of
Einstein equations for a rather wide class of Lagrangians.

We emphasize that, incidentally, our result shows also that the
metric--affine approach is not equivalent to
the purely metric
one for non--linear
Lagrangians. In fact, for Lagrangians which
depend non--linearly on the scalar curvature of a
metric, it is known
that higher derivatives effects entail the
 existence of a further
scalar field dynamically interacting with the graviton
(see  \cite{UDW,DTN,Ste,Whi,MFF1,Mae,MFF2} and
 ref.s quoted therein),
while here
only a cosmological constant appears.
The relations between these two formalisms and with
conformal invariance are discussed elsewhere \cite{FMV1}.

Although for our present purposes we are
interested mainly in
4--dimensional space--times, we shall discuss
field equations in the
arbitrary dimension $n \geq 2$, in view of
full generality
and applications to 2--dimensional gravity
theories. As is well known,
in fact, in two dimensions Einstein equations
do no longer exist
but can be appropriately replaced by other equations involving
the curvature (see, e.g., \cite{Jac}). Our
results
will in fact provide a new ``universal'' equation also for $n=2$,
which states that the scalar curvature of the metric
 and the connection
is a constant. This universal equation has a
purely geometrical
origin and does not require the introduction of
 extra fields having
no geometrical origin.
Its important relations with a new topological
theory of gravity based on a purely affine Lagrangian
are also discussed
elsewhere \cite{FFV1}.

\section{The Universal Field Equations}

Let us consider an action of the form

$$
S(\Gamma, g)=\int_M L(R)\sqrt g \ d^nx \eqno        (1)
$$

\noindent on a n--dimensional manifold $M$ endowed with a metric
$g_{\mu \nu}$ and a torsionless
connection $\Gamma_{\mu \nu}^{\sigma}$.
We use the following standard notation:

$$
R_{\mu \nu \sigma}^{\lambda}=\partial_{\nu} \Gamma_{\mu
\sigma}^{\lambda} - \partial_{\sigma}\Gamma_{\mu
 \nu}^{\lambda} + \Gamma_{\alpha\nu}^{\lambda} \Gamma_{\mu
\sigma}^{\alpha} - \Gamma_{\alpha \sigma}^{\lambda}
\Gamma_{\mu \nu}^{\alpha}
$$

$$
R_{\mu \sigma} = R_{\mu \sigma}(\Gamma)=R_{\mu \nu \sigma}^{\nu} \ ,
R=R(\Gamma,g)=g^{\mu \sigma}R_{\mu \sigma}(\Gamma) \ ,
\alpha, \mu, \nu,...=1,...,n
$$

\noindent The Lagrangian $L$ is a given function of
 one real variable,
which we assume to be real analytic on its domain of definition.

The Euler--Lagrange equations for the action (1) with respect to
independent variations of
$g$ and $\Gamma$
can be written in the following form

$$
L^{\prime} (R)R_{(\mu \nu )}(\Gamma)-{1\over 2}L(R)g_{\mu
\nu}=0\eqno (2)
$$

$$
\nabla_{\alpha}(L^{\prime} (R)\sqrt g \ g^{\mu \nu})=0\eqno (3)
$$

\noindent where $ \nabla_{\alpha}$ is the covariant
derivative with respect
to $\Gamma$, $R_{(\mu\nu)}$ denotes the symmetric part of the
Ricci tensor of $\Gamma$ and we assume $n\geq 2$.
In fact, variation of the action (1) with respect to $\Gamma$
gives the equation:

$$
\nabla_{\alpha}(L^{\prime}(R)\sqrt g \ g^{\mu\nu}) -
\nabla_{\rho}L^{\prime}(R)\sqrt g
\ g^{\rho (\mu}\delta^{\nu )}_\alpha=0
$$

\noindent which, in any dimension $n \geq 2$, reduces to (3) by
taking a trace.

Quadratic Lagrangians have been investigated by
 several authors, mainly
in the four--dimensional
case and also in the presence of matter (see, e.g.,
\cite{Step,Hig,Yan,Heh,Fai,Sha}
 and ref.s quoted therein); their
generalizations in the presence of torsion have also
received a lot of attention
in the literature (see \cite{Heh,Nee,Nev,Sez,KV,Ros}
J and ref.s
quoted therein). General non--linear Lagrangians
 of the form (1) in dimension
$n=4$ and in the presence of matter have been independently
investigated in \cite{Ham} from a  different
 viewpoint, with the aim
of discussing conservation laws and the validity
of the strong equivalence principle. The results
 of \cite{Ham} concerning
field equations are in
agreement with ours and are in fact a particular case of ours;
it should be mentioned that earlier results were previously
found for quadratic Lagrangians in \cite{Step,Hig,Sha}.

In this paper we
discuss from a different and apparently new viewpoint
the general structure of
equations (2) and (3) above, in an arbitrary dimension $n \geq 2$
and for an arbitrary analytic function $L(R)$. We recognize
and emphasize a ``universal property'' of this whole family
of Lagrangians which, to our knowledge, has been never clearly
formulated in the existing literature. We prove, in fact, that the
family generates ``universal equations'', independently from the
particular Lagrangian chosen, and that in dimension $n>2$ these
equations coincide with Einstein equations in the generic case.
We further discuss
the ``universal equations'' in dimension
$n=2$ and for all "degenerate" cases, which include the constant
scalar curvature equation and conformally invariant equations
for conformally invariant Lagrangians.

To solve equations (2)--(3) we proceed as follows.
First of all, by taking the trace of eq. (2) one obtains the following
equation:

$$
L^{\prime} (R)R-{n\over2}L(R)=0 \eqno (4)
$$

\noindent We shall then distinguish the following three mutually exclusive
cases: (i) eq. (4) has no real solutions; (ii) eq. (4) has real solutions;
(iii) eq. (4) is identically satisfied. Their
 discussion proceeds as follows:

\noindent {\bf (Case 1)} If equation (4) has no real solutions, then
also the system (2)--(3) has no real solution.

\noindent {\bf (Case 2)} Let us now suppose
that eq. (4) is not identically
satisfied and has at
least one real solution. In this case, since
 analytic functions can have
at most a countable set of zeroes on the real line, eq. (4)
can have no more than a countable set of solutions $R=c_i (i=1,2,\ldots)$,
where $c_i$ are constants.
Consider then any solution

$$
R=c_i \eqno (5)
$$

\noindent of eq. (4). We have again two
 possibilities, depending on
the value of the first
 derivative $L^{\prime}(c_i)$ at the point $R = c_i$:

\noindent {\bf (Subcase 2.1)} Let us assume that
$L^{\prime}(c_i) \neq 0$. Then eq. (3) takes the form

$$
\nabla_{\alpha}(\sqrt g \ g^{\mu \nu})=0 \eqno (6)
$$

\noindent while equation (2) reduces to

$$
R_{(\mu \nu)}(\Gamma)=\Lambda(c_i) g_{\mu \nu}   \eqno (7)
$$

\noindent where a cosmological constant $\Lambda =
\Lambda (c_i)$ arises
according to:

$$
\Lambda = \Lambda(c_i) =L(c_i)/2L^{\prime}(c_i)=
c_i/n  \eqno (8)
$$

We shall discuss separately the subcases $n>2$ and $n=2$.

\noindent {\bf (Subcase 2.1.1)}

For $n>2$ and any metric $g_{\mu \nu}$
the general solution of eq. (6) is the Levi--Civita
connection  of the given metric $g$,
so that the Ricci tensor $R_{\mu \nu}(\Gamma)$
is automatically symmetric and in fact identical to
the Ricci tensor
$R_{\mu \nu}(g)$ of the metric $g$ itself. Accordingly, the
general solution
of the system (3)--(4) is given by

$$
\Gamma_{\mu \nu}^{\sigma}=\Gamma_{\mu \nu}^{\sigma}(g)=
{1\over2}g^{\sigma \alpha}(\partial_{\mu}g_{\nu \alpha}
+\partial_{\nu}g_{\mu \alpha}-\partial_{\alpha}g_{\mu \nu})
\eqno (9)
$$

\noindent where the metric $g_{\mu\nu}$ is any solution of the
equations

$$
R_{\mu \nu}(g)=\Lambda(c_i) g_{\mu \nu}   \eqno (10)
$$

\noindent with cosmological constant $\Lambda$ given by (8). This constant may
vanish if and only if $c_i=0$ is one of the solutions of eq. (4).

\noindent {\bf (Subase 2.1.2)}
In the case $n=2$, instead, equation (6) allows a further degree
of freedom (related with conformal invariance) and it has in fact
the following general solution:

$$
\Gamma_{\mu \nu}^{\sigma}=W_{\mu \nu}^{\sigma}(g,B)=
{1\over2}g^{\sigma \alpha}(\partial_{\mu}g_{\nu \alpha}
+\partial_{\nu}g_{\mu \alpha}-\partial_{\alpha}g_{\mu \nu})
+{1\over 2}(\delta^\mu_\nu B_\sigma + \delta^\mu_\sigma B_\nu
            - g_{\mu\nu}B^{\alpha})
\eqno (11)
$$

\noindent where $B^\alpha$ is an arbitrary
 vectorfield. This is due to
the fact that only in dimension $n=2$ eq. (6) cannot be reduced
to $\nabla_{\alpha}g^{\mu \nu}=0$ (since only in two dimensions
it does not imply that the Riemannian volume
element of $g$ is covariantly
constant along $\Gamma$). A connection $W(g,B)$
having the form (11) will be hereafter called
a {\it Weyl connection} (\cite{Wey}).
Using eq. (11), from the definition
 of $R_{(\mu\nu)}(\Gamma)$ one has:

$$
R_{(\mu\nu)}\equiv {1\over 2}(R(g) - D_\alpha B^\alpha)g_{\mu\nu}
\eqno (12)
$$

\noindent so that eq. (7)
reduces to the following scalar equation

$$
R(g) - D_\alpha B^\alpha = 2 \Lambda
\eqno (13)
$$

\noindent where $D_\alpha$ denotes the covariant
derivative with respect to
the metric $g_{\mu\nu}$. Equation (13) is
 the ``universal'' equation
for 2--dimensional space--times; it replaces
 Einstein equations,
which are the ``universal equations'' in dimension
$n>2$. Equation (13) is in fact the equation of
 constant scalar curvature
for the metric g and the Weyl connection (11), because from (12)
one has:

$$
{\cal R}(g,B) = g^{\mu\nu} R_{\mu\nu}(W(g,B)) =
 R(g) - D_\alpha B^\alpha
\eqno (14)
$$

We remark that eq. (13) has always infinitely
many local solutions, but it might have no
 global analytic solution
(depending on the topology of the 2-dimensional manifold $M$).

\noindent {\bf (Subcase 2.2)} Suppose now $L^{\prime}(c_i)=0$.
Then eq. (4) implies that also $L(c_i)=0$, i.e., $c_i$ is
a zero of order at
least two of $L(R)$. In this case, eq.s (2)--(3) are identically
satisfied and the only relation
 between $g$ and $\Gamma$ is contained
in the following equation

$$
R(g,\Gamma) = c_i     \eqno (15)
$$

This equation represents a genuine dynamical relation between the
metric and the connection of $M$, although it is not enough to single
out a connection $\Gamma$ for any given metric $g$ (as it happened,
instead, in subcase (2.a) above, where $\Gamma$ turns out to be the
Levi--Civita connection of $g$). In fact, defining a tensorfield
$\Delta^{\lambda}_{\mu\nu}$ by:

$$
\Delta^{\lambda}_{\mu\nu} \equiv \Delta^{\lambda}_{\mu\nu}(g,\Gamma) =
\Gamma^{\lambda}_{\mu\nu} - \Gamma^{\lambda}_{\mu\nu}(g)
\eqno (16)
$$

\noindent equation (14) can be turned into a
 quasi--linear first--order PDE for the
unknown $\Delta$, having the term $R(g) - c_i$ as a source. The
space of solutions of this last equation, as functions of the metric
$g$ together with its first derivatives and a number of auxiliary fields,
has a complicated structure. In any case, it is easy
to see that this space
contains as a subspace the space of all couples $(g,\Gamma)$
satisfying eq. (7) for $\Lambda = {c_i\over n}$.

\noindent {\bf (Case 3)} We consider now the case
in which eq. (4) is
identically satisfied. Under this hypothesis
the Lagrangian is
proportional to:

$$
L(R)=|R|^{n/2}\eqno (17)
$$

\noindent This Lagrangian is analytic
everywhere on the real line, except at the point $R=0$ (which is
singular), unless $n=4k$ for some positive integer $k$. We
shall then
consider for simplicity the case $R \geq 0$ (analogous
 results will
be valid for $R \leq 0$ and they may extend
across $R=0$ at least
if $n=4k$). We stress that for $n=4$ this is exactly
 the quadratic
Lagrangian $L(R)=R^2$ considered in \cite{Step,Hig} so that our
results will suitably complete and extend the
earlier discussion
appearing therein.
\par

In this case equations (2) and (3) read as follows:

$$
 R^{n-2\over 2}
    ( R_{(\mu\nu)} - {1\over n} R g_{\mu\nu} ) = 0
\eqno (18)
$$

$$
\nabla_\alpha (R^{n-2\over 2}
    \sqrt g g^{\mu\nu} ) = 0 \eqno (19)
$$

Notice first of all that, under conformal transformations

$$
\tilde g_{\mu\nu} = e^{\omega}g_{\mu\nu} \    , \   \tilde \Gamma = \Gamma
\eqno (20)
$$

\noindent setting

$$
R = g^{\mu\nu}R_{\mu\nu}(\Gamma) \ ,
\tilde R = \tilde g ^{\mu\nu}R_{\mu\nu}(\tilde \Gamma) =
\tilde g ^{\mu\nu}R_{\mu\nu}(\Gamma)
\eqno (21)
$$

\noindent one has

$$
\begin{array}{rclr}
\hskip4cm\tilde R &=&  e^{-\omega}R \ , &(22.a)\\
\tilde R \tilde g _{\mu\nu} &=& Rg_{\mu\nu} \ ,
&(22.b)\\
\tilde R ^{n/2} \sqrt{\tilde g} &=& R^{n/2}\sqrt g \ , &(22.c)\\
\tilde R ^{n-2\over 2} \sqrt{\tilde g}\tilde g ^{\mu\nu} &=&
R^{n-2\over 2} \sqrt g g^{\mu\nu} \ ,
&\hskip3.4cm (22.d)
\end{array}
$$

\noindent Therefore, the action

$$
S(g,\Gamma) = \int_M |R|^{n/2} \sqrt g d^nx
$$

\noindent as well as equations (18) and (19) are
invariant under the
transformation (20), i.e. $S(\tilde g,\tilde \Gamma) = S(g,\Gamma)$.

\noindent Also in this case we shall consider
 separately the two
cases $n>2$ and $n=2$.

\noindent {\bf (Subcase 3.1)}

If $n>2$ we have two possibilities. If $R=0$, then
 we have only the
equation

$$
R(g,\Gamma) = 0
\eqno (23)
$$

\noindent whose discussion proceeds as for
 eq. (15) above. When $R>0$ there is
instead an additional conformal degree of
freedom, as noticed earlier
in \cite{Hig}J
(and later exploited in \cite{Fer}). In this case, in fact,
eq.s (18) and (19) reduce to the following:

$$
 R_{(\mu\nu)} - {1\over n} R g_{\mu\nu} = 0
\eqno (24)
$$

$$
\nabla_\alpha (R^{n-2\over 2}
    \sqrt g g^{\mu\nu} ) = 0 \eqno (25)
$$

The following proposition is then true:

\newtheorem{proposition}{Proposition}
\begin{proposition}
{\bf(i)} If $h_{\mu\nu}$ is a solution of the equation

$$
R_{\mu\nu}(h) = h_{\mu\nu} \ ,
\eqno (26)
$$

where $R_{\mu\nu}(h)$ is the Ricci tensor of the metric $h_{\mu\nu}$,
then the pair $(g,\Gamma)$, where

$$
\Gamma = \Gamma_{LC}(h)
\eqno (27)
$$

$$
g_{\mu\nu}J= e^{-\omega}h_{\mu\nu}
\eqno (28)
$$

is a solution of eq.s (24) and (25) for any function $\omega$
(here, $\Gamma_{LC}$ is the Levi--Civita connection of $h$).

\noindent{\bf (ii)} If $(g,\Gamma)$ is a solution of eq.s (24) and (25), then
they have to satisfy the relations

$$
R_{\mu\nu}(a) - {1\over n} a_{\mu\nu} = 0
\eqno (29)
$$

$$
\Gamma = \Gamma_{LC}(a)
\eqno (30)
$$

where

$$
a_{\mu\nu}J= R(g,\Gamma)g_{\mu\nu}
\eqno (31)
$$

\end{proposition}

\noindent{\it Proof.} From eq. (27) one has

$$
R_{(\mu\nu)}(\Gamma)=R_{\mu\nu}(h)
\eqno (32)
$$

\noindent We shall now use eq.s (20)--(22) for $\tilde g = h$ to find:

$$
\tilde R = h^{\mu\nu}h_{\mu\nu} = n
\eqno (33)
$$

\noindent Using finally eq. (22.b),(26),(32) and (33) one gets:

$$
R_{(\mu\nu)}(\Gamma) - {1\over n} R g_{\mu\nu} =
R_{\mu\nu}(h) - {1\over n} \tilde R \tilde g _{\mu\nu} =
h_{\mu\nu} - {1\over n}n h_{\mu\nu} = 0
$$

which shows that (24) holds for the
 couple $(g,\Gamma)$. To prove that the
couple satisfies also eq. (25) we
 use (22.d) and (33), to obtain:

$$
\nabla_\alpha (R^{n-2\over 2} \sqrt g g^{\mu\nu} ) =
D_\alpha (\tilde R ^{n-2\over 2}
 \sqrt {\tilde g} \tilde g ^{\mu\nu} ) =
n^{n-2\over 2} D_\alpha (\sqrt h h^{\mu\nu} ) = 0
$$

\noindent where $D_\alpha$ denotes the covariant
 derivative with
respect to the Levi--Civita connection of $h$.
This proves (i). We proceed now to prove (ii). Using
 formulae (20)--(22)
for $\tilde g _{\mu\nu}J= a_{\mu\nu}
 \ , e^{\omega} = R(g,\Gamma)$ one
finds $\tilde R = 1$. From (22.d) one gets then:

$$
R^{n-2\over 2} \sqrt g g^{\mu\nu} = \sqrt a a^{\mu\nu}
$$

\noindent so that eq. (25) now reads as follows:

$$
\nabla_\alpha (\sqrt a a^{\mu\nu}) = O
\eqno (34)
$$

\noindent which gives immediately eq. (30). From this
 it follows in
turn that $R_{(\mu\nu)}(\Gamma) = R_{\mu\nu}(a)$. Using
then $\tilde R = 1$
together with (22.b) gives $R g_{\mu\nu} = a_{\mu\nu}$. Therefore, eq.
(29) follows from (24). This completes our proof. \ \ \ {\bf (Q.E.D)}

The above proposition can be re--phrased as follows.
If we define a new metric $h_{\mu\nu}$ by setting:

$$
\sqrt{h} h^{\mu\nu} = R^{n-2\over 2}\sqrt{g} g^{\mu\nu}
\eqno (35)
$$

\noindent then eq.s (19) and (22) imply
that $\Gamma$ is the Levi--Civita
connection of the new metric $h$ and eq. (18) reduces to

$$
R_{\mu\nu}(h) - {1\over n}h_{\mu\nu} = 0
\eqno (36)
$$

\noindent This, in turn, leads to a constant scalar
curvature for the new
 metric $h$ (in fact, it is $R(h) = 1$) and $M$ will
be an Einstein manifold with respect to the new metric $h$.
According to the earlier discussion
of \cite{Hig,Fer} we can also restate the
result as follows: if $(g,\Gamma)$ is a
 solution of eq.s (18) and (19),
then there exists a scalar field $\psi$ such
that the conformally related
metric $\psi g_{\mu\nu}$ satisfies Einstein
equations and the connection
$\Gamma$ is the Levi--Civita connection
 of $\psi g_{\mu\nu}$; moreover,
the scalar curvature of the original
 metric $g$ and $\Gamma$ equals $\psi$.
The origin of this extra scalar field $\psi$ is
 discussed elsewhere (\cite{FMV1})
in the framework of Legendre transformation
for metric--affine theories.

\noindent {\bf (Subcase 3.2)}

If $n=2$ then equations (18) and (19) simplify to

$$
R_{(\mu\nu)}(\Gamma) - {1\over 2} R(g,\Gamma) g_{\mu\nu} = 0
\eqno(37)
$$

$$
\nabla_\alpha ( \sqrt{g} g^{\mu\nu} ) = 0
\eqno(38)
$$

Because of (12) and (14), the general solution of
 equation (38) is
again represented by (11), and this in turn implies
that equation
(37) is identically satisfied. Therefore, the general
solution of
equations (37)--(38) is given by a
 pair $(g,W(g,B))$ where $g$ is
an arbitrary metric and $W(g,B)$ is a
Weyl connection, determined by the same metric and an
 arbitrary vectorfield
$B$. Equation (13) is now replaced by

$$
R(g) - D_\alpha B^\alpha = \Lambda (x)
\eqno (39)
$$

\noindent where now $\Lambda (x)$ is an arbitrary function.

\vskip1truecm

We can then summarize our results above in the following:

\newtheorem{theorem}{Theorem}
\begin{theorem}

Let $L(R)$ be an arbitrary
analytic Lagrangian in a $n$--dimensional
manifold $M$, which depends on the scalar
 curvature $R(g,\Gamma)$ of a metric $g$
and a torsionless connection $\Gamma$. The dynamical
 behaviour of $(g,\Gamma)$
is governed by the equation

$$ L^{\prime} (R)R-{n\over2}L(R)=0 \eqno (*) $$

\noindent Then, either one of the  following holds:
\begin{description}
\item[(1)] Equation (*) has no real solutions.
\item[(2)] Equation (*) has a discrete set of real
 solutions $R=c_i,i=1,2,\ldots$
\item[(3)] Equation (*) is identically satisfied; in
 this case the Lagrangian
is proportional to to the power $R^{n/2}$.
\end{description}

\noindent Accordingly, either one of the following holds:
\begin{description}

\item[(1)] If eq. (*) has no real solutions than there
 are no consistent
field equations.
\item[(2.1)] If $n>2$ and $R=c_i$ is a solution
of eq. (*) such that
$L^{\prime}(c_i) \neq 0$ then $\Gamma$ is the Levi--Civita
connection of $g$ and $g$ satisfies Einstein equations with
 cosmological constant
$\Lambda = c_i/n$.
\item[(2.2)] If $n=2$ and $R=c_i$ is a solution
 of eq. (*) such that
$L^{\prime}(c_i) \neq 0$ then $g$ is an arbitary metric
 and  $\Gamma$ is
the Weyl connection
$W(g,B)$ generated by the Levi--Civita
connection of $g$ together with an arbitrary
 vectorfield $B$. The pair $(g,B)$
satisfies the equation ${\cal R}(g,B) = R(g,W(g,B)) = c_i$.
\item[(2.3)] If $n \geq 2$ and $R=c_i$ is a solution of
 eq. (*) such that
$L^{\prime}(c_i) = 0$ then the only dynamical relation
 between $\Gamma$ and $g$
tells that $R(g,\Gamma) = c_i$.
\item[(3.1)] If $n>2$ then either $R(g,\Gamma)=0$ is the
 only dynamical relation
between $\Gamma$ and $g$ or (if $R \neq 0$) the following holds:
if $(g,\Gamma)$ is a solution
then there exists a scalar field $\psi$ such that the
 conformally related
metric $\psi g_{\mu\nu}$ satisfies Einstein equations and
 the connection
$\Gamma$ is the Levi--Civita connection
 of $\psi g_{\mu\nu}$; moreover,
the scalar curvature of the original
 metric $g$ and $\Gamma$ equals $\psi$.
\item[(3.2)] If $n=2$ then $g$ is an arbitary metric
 and  $\Gamma$ is the Weyl
connection $W(g,B)$ generated by the Levi--Civita
 connection of $g$ and
by an arbitrary vectorfield $B$.

\end{description}
\end{theorem}

Let us make some further comments on
the ``exceptional'' Lagrangian (17),
i.e. $L(R)=a|R|^{n \over 2}$ (which
is degenerate in the appropriate dimension $n$). We
first remark that
this Lagrangian is in fact invariant under conformal
 rescalings of the metric $g$,
with $\Gamma$ fixed (further comments on this may
 be found in \cite{FMV1}). This
is particularly relevant for 4--dimensional
space--times, where the ``exceptional case'' is just
 the quadratic
Lagrangian $L(R)=R^2$; this case was already
 considered in \cite{Hig}, where
it was argued that it always leads to Einstein
 equations (for a conformal
family of metrics). It turns out that this is
 fact true only for $R \neq 0$,
as a particular case of our general discussion.
 We remark, however, that
the case $R(g,\Gamma)=0$, which was a priori
 excluded in \cite{Hig}, has in fact
a great relevance as it was discussed above; it
 does not lead to Einstein
equations but to a larger space of solutions, contradicting
 the conclusions of
\cite{Hig}.

As a final remark, we notice that if we add to the action (1)
a matter Lagrangian $L_{\rm mat}(g, \psi,\partial\psi)$ describing the
minimal coupling of the metric $g$
with external matterfields
$\psi$, the eq. (3) remains unchanged, while instead of
equation (2) one finds:

$$
L^{\prime} (R)R_{(\mu \nu )}(\Gamma)-{1\over2}L(R)g_{\mu
\nu}= T_{\mu\nu}
\eqno (40)
$$

\noindent where $T_{\mu\nu} \equiv \delta L_{\rm mat}/\delta g_{\mu\nu}$
is the energy--momentum tensor of matter. Taking the trace of
equation (40) one gets then:

$$
L^{\prime} (R)R-{n\over2}L(R)=T \eqno (41)
$$

\noindent where $T=g^{\mu\nu}T_{\mu\nu}$. If $T$ is zero or constant
the same considerations as above lead therefore to analogous
conclusions about the universality of Einstein equations
(or of their counterpart (13) for special cases). More general
interactions of the connection $\Gamma$ with matter will
be considered elsewhere \cite{FFV2}.

\section{Examples}

As an example, we shall consider the space of all Lagrangians
 of the form

$$
L(R)=aR^2+bR+c \eqno (42)
$$

\noindent which is identified to the three--dimensional
 space $\R^3$
with parameters $(a,b,c)$. If $n=4$, for any
 point $(a,b,c) \in \R^3$
with $b\not=0$ and $b^2-4ac\not=0$ one gets Einstein
 equations (7)
with $\Lambda=-c/2b$.
Therefore, in this case the space $E\subset \R^3$ of Lagrangians
leading to Einstein equations contains all points of $\R^3$ with
$b\not=0$ and $b^2-4ac\not=0$. On the contrary, on the surface
$\{b^2-4ac=0,b\not=0\}$,
instead, we have not Einstein equations and the only dynamical
relation between $g$ and $\Gamma$ is given by the
equation

$$
R(\Gamma,g)=-2c/b \eqno (43)
$$

\noindent This surface is therefore a
``bifurcation surface'' in the space of all Lagrangians; in fact,
when coupling constants in (42) let $L(R)$ tend to the surface,
the functional space of solutions $(g,\Gamma)$ enlarges from
the space of pairs $(g,\Gamma(g))$ satisfying Einstein equations
to the larger space of all pairs $(g,\Gamma)$ satisfying
eq. (43).
The line $\{b=c=0,a\not=0\}$ corresponds to the
exceptional case $L(R)=aR^2$. In this case there are
 solutions $(g,\Gamma)$ of
eqs. (2)--(3) which are described by Einstein equations
 and there are also
pairs $(g,\Gamma)$ with the only restriction $R(g,\Gamma)=0$.
Finally, we mention that for $b=0$ and $c\not=0$
one gets an inconsistent system of equations.

Another interesting example is

$$
L(R) = R + aR^k
\eqno (44)
$$

This Lagrangian in $n$--dimensional space--time ($n>2$) gives
Einstein equations for any $a$ and any $k=2,3,\ldots$ (if
$k \neq n/2$). In particular, in 4--dimensional space--times
if $k$ is odd
and $a \leq 0$ the corresponding Einstein equations are
$R_{\mu\nu}(g)=0$. Notice here also that the Lagrangian
$L(R) = R + aR^2$ gives Einstein equations $R_{\mu\nu}(g)=0$
for any $a$.

\section{Conclusions}

We have shown that non--linear Lagrangians depending
 on the scalar
curvature $R(g,\Gamma)$ always lead to ``universal'' equations as
Euler--Lagrange equations, unless it does not lead to any
 consistent equation
at all. Our results were obtained under the explicit
 hypothesis that
the Lagrangian $L$ is an analytic function. However, since they
depend on solutions of eq. (4), ours results above trivially
 extend to
all $C^2$ Lagrangians such that eq. (4) has a discrete set
 of solutions
in the domain of definition of $L$.
In dimension $n>2$ these universal equations are either Einstein
equations with a cosmological constant, in a generic
 case, or the constant
scalar curvature equations $R(g,\Gamma) = {\it constant}$, in
degenerate cases. A notable exception is the case
$L(R)=aR^{n/2}$, in which an extra conformal degree of
freedom appears.
In dimension $n=2$ conformal invariance entails
instead that the
connection is a Weyl connection, in which a
 vectorfield $B$ appear
alongwith the Levi--Civita connection of a metric; the universal
equations still involve $R(g,\Gamma)$.

{}From a functional viewpoint, we argue
that ``most'' (analytic) Lagrangians
depending on the scalar curvature $R(g,\Gamma)$ either
 lead to Einstein
equations or do not give any
consistent equations at all, provided we endow the space of
 all (analytic)
Lagrangians  with a reasonable topology (i.e., the Lagrangians
 considered
above represent a ``generic'' case). Out of these generic points
we have in fact a bifurcation and Einstein equations are replaced
 by other universal
equations which state that $R(g,\Gamma)$ is a
 constant. However, from a ``practical
viewpoint'', the physical parameters
in a Lagrangian
are known only approximately. This means that we are not
 dealing with a uniquely given
Lagrangian with fixed coupling constants, but rather with
 a family of Lagrangians,
and even small perturbations may destroy the structure of
 equations at bifurcation
points.
It would be therefore interesting to investigate in greater
 detail the geometry
of the functional space
of all Lagrangians leading to Einstein equations. This should
 have relevance
also to the problem of quantization of the gravitational
 field by means
of non--linear Lagrangians and also when viewing non--linear
 gravity as a
low--energy limit of string theory.

We also point out that in this paper we have
 proved ``universality'' for field
equations in a large class of non--linear Lagrangians, but
we have not addressed the important problem
 of ``universality'' of physical
observables (like, e.g., conserved quantities). It should
 be therefore
important to investigate the
role of our results above in connection with
energy and conservation laws (e.g., by means of
 the Poincar\'{e}--Cartan formalism),
in order to see to what extent universality holds
 or not also at this
level. In particular,
it will be interesting to compare the notion of energy
one obtains directly
from Einstein equations with the energy one calculates
starting from
the Lagrangian. We aim to discuss this problem in
 a future investigation.

It should be stressed that, as we mentioned above, in the purely
 metric formalism an
additional scalar field
appears when dealing with non--linear
 Lagrangians $L(R)$ (see \cite{MFF1,MFF2}),
while here we obtain only the standard Einstein
equations and a cosmological
constant (unless $L(R)=aR^{n/2}$). Further comments on
 this difference of behaviour
are discussed elsewhere \cite{FMV1} through the method of
 Legendre transformation
and related with conformal invariance under rescalings
 of the metric.

A difference between the metric and vierbein--connection formalisms
was discussed by Witten in \cite{Wit}, where it was shown
 that the application of the
vierbein--connection formalism to 3--dimensional gravity allows to
prove its solvability and renormalizability. As it follows from our
discussion, by using  the metric--affine formalism in any dimension
$n\geq 3$ one can  add to the Hilbert Lagrangian any counterterms
depending on the scalar curvature without changing the physical
content of the theory (i.e., the Einstein equations, if one does not
use ``fine tuning'', i.e. one assume to be in the generic
 case). Only the
cosmological constant will in fact be renormalized, according to (8).

It would also be interesting to study under which
 conditions the approach
presented in this note
can be extended to more general classes of
 Lagrangians leading again to
Einstein equations, including a dependence
on invariants more complicated than the scalar
 curvature (e.g., the square
of the Ricci tensor). This is currently under
 investigation (\cite{FFV2,FMV2}).

\section{Acknowledgements}

We are deeply grateful to G. Magnano for his useful
 remarks. One of us (I.V.)
gratefully acknowledges
the hospitality of the Institute of Mathematical
 Physics ``J.--L. Lagrange''
of the University of Torino and the continuative
 support of G.N.F.M. of Italian
C.N.R. This work is sponsored
 by G.N.F.M., M.U.R.S.T. (40\% \ Proj.
``Metodi Geometrici e Probabilistici in
 Fisica Matematica''); one of us
(M. Ferraris) acknowledges also
 support from I.N.F.N.

\end{document}